\documentclass[]{spie}  

\usepackage{amsmath,amsfonts,amssymb}
\usepackage{graphicx}
\usepackage[colorlinks=true, allcolors=blue]{hyperref}

\title{Preparing the NIRSpec/JWST science data calibration: from ground testing to sky}

\author[a]{Catarina Alves de Oliveira}
\author[a]{Stephan M. Birkmann}
\author[a]{Torsten B\"oker}
\author[b]{Pierre Ferruit}
\author[b]{Giovanna Giardino}
\author[a]{Nora L\"utzgendorf}
\author[a]{Elena Puga}
\author[a]{Tim Rawle}
\author[a]{Marco Sirianni}
\author[a]{Maurice te Plate}
\affil[a]{European Space Agency, c/o STScI, 3700 San Martin Drive, Baltimore, USA}
\affil[b]{European Space Agency, ESTEC, Keplerlaan 1, 2200AG Noordwijk, The Netherlands}

\authorinfo{Further author information:\\C.A.d.O.: E-mail: catarina.alves@esa.int}

\begin{document} 
\maketitle

\begin{abstract}
The Near-Infrared Spectrograph (NIRSpec) is one of four instruments aboard the James Webb Space Telescope (JWST). NIRSpec is developed by ESA with AIRBUS Defence \& Space as prime contractor. The calibration of its various observing modes is a fundamental step to achieve the mission science goals and provide users with the best quality data from early on in the mission. Extensive testing of NIRSpec on the ground, aided by a detailed model of the instrument, allow us to derive initial corrections for the foreseeable calibrations. We present a snapshot of the current calibration scheme that will be revisited once JWST is in orbit.  
\end{abstract}


\section{INTRODUCTION}
\label{sec:intro}  
The Near-Infrared Spectrograph (NIRSpec) is one of the four instruments aboard the James Webb Space Telescope (JWST). NIRSpec is developed by ESA with AIRBUS Defence \& Space as prime contractor. The calibration of the NIRSpec instrument and its science data is a fundamental step to achieve the mission science goals and provide users with the best quality data from early on in the mission. NIRSpec is a versatile spectrograph equipped with an integral field unit (IFU), five fixed slits (FS) and a micro-shutter array (MSA) that provide JWST with 3D, high-contrast, and multi-object spectroscopic (MOS) capabilities, respectively. The combination of NIRSpec's filters and dispersive elements covers the 0.6 to 5.3 micron spectral range at three different spectral resolutions, with a total of nine different configurations. In addition, a mirror element provides an imaging capability, used for on-board target acquisition and calibration.

This impressive versatility makes NIRSpec a very powerful and attractive instrument from the scientific point of view, but turns into a challenge when it comes to calibration. The situation is made even more complex by the fact that the disperser wheel sitting at the heart of the instrument has a limited angular positioning repeatability. Each time it comes back to a given disperser or to the mirror, it stops at a slightly different position. This has the consequence of changing the location of spectra or a target acquisition image on the detector for all exposures, precluding a static calibration of the wavelength or spatial solution for each configuration. 

In this contribution we present a high-level summary of the solutions found to calibrate the NIRSpec instrument and its science data. A status update on the results from the latest ground-test campaign of the instrument can be found elsewhere in these proceedings\cite{NIRSpecStatus}. 

\section{Data: ground-test campaigns, simulations, and plans for commissioning}
\label{sec:data}  
The NIRSpec calibration plan has been designed to ensure that it uses the best possible data that can be obtained for each subsystem and tackles the three main parts of the optical train: detectors, spectrograph, and FORE optics (including the telescope). The present calibration is based on data acquired in the laboratory at component level, and during ground testing involving the instrument either alone or with the telescope and the other instruments. It uses exposures obtained with internal and external test lamps, as well as data produced by the instrument performance simulator. Once in orbit, complementary data will be incorporated using NIRSpec internal lamps and, most importantly, on-sky observations. The current snapshot of the calibration scheme is therefore a hybrid between these elements and datasets that will ultimately be evaluated and revisited once in orbit. In this section, we provide a short description of the data available. 

\subsection{On-the-ground test campaigns}
The NIRSpec instrument has undergone a series of test campaigns under cryogenic vacuum conditions. First, two instrument-level testing cycles took place at IABG facilities in Ottobrunn, Germany, in 2011 and 2013. NIRSpec was enclosed in a shroud cooled by gaseous Helium and inside a dedicated chamber together with optical ground-support equipment that included calibration light sources as well as a cryo-mechanism allowing to switch between flat-field illumination and a grid of pinholes. Data acquired in these campaigns demonstrated the expected performance of the NIRSpec instrument\cite{2014SPIE.9143E..08B,2012SPIE.8442E..3EB}, and served as the baseline dataset to develop the calibration plan for NIRSpec. The instrument was then delivered to NASA in 2013, where it was installed on the Integrated Science Instrument Module (ISIM) together with the other JWST science instruments. Two environmental and cryogenic test campaigns of ISIM followed (cryo-vacuum tests CV2 and CV3) in 2014 and 2016 at the Goddard Space Flight Center (GSFC). Between those, the NIRSpec instrument was removed from ISIM and refurbished with two new subsystems, the MSA and the FPA, where parts that had suffered degradation were replaced \cite{2016SPIE.9904E..0DT}. These campaigns, and in particular the second with NIRSpec in its flight configuration, were crucial to verify the health and performance of the instrument after vibration and acoustic testing at ISIM level, and also gather the necessary data to update and expand the calibration of the instrument \cite{2016SPIE.9904E..0BB}. The next milestone was the integration of ISIM with the Optical Telescope Element (OTE), which together form the Optical Telescope Element and Integrated Science (OTIS). OTIS underwent environmental testing (vibe and acoustic) in fall 2016, and after being shipped to Johnson Space Center (JSC) it was subjected to an extensive thermal vacuum campaign in the summer of 2017. The tests conducted in this last test campaign were used to verify the health, integrity, and performance of NIRSpec, and also carried out as much as possible in a flight-like manner, in accordance with the procedures planned for in-orbit commissioning \cite{2016SPIE.9904E..44B}.

For calibration purposes, we also make use of data acquired with tests performed at single-component level, since those measurements are no longer possible in the integrated instrument. Examples of such datasets are the spectral response of NIRSpec's internal lamps or the filter's throughput curves. Another example are the performance characterization tests carried out on NIRSpec's detectors in the Goddard Detector Characterization Laboratory (DCL) \cite{2014PASP..126..739R}. Only data acquired during these tests are suitable to derive the contribution of the detector response to the flat-field corrections, since after the integration of the focal plane array (FPA) into NIRSpec, the presence of the MSA in the optical path no longer permits a uniform illumination of the entire area of the two Sensor Chip Assemblies (SCAs).
 
\subsection{Simulations}
Data collected in ground-test campaigns has been fundamental in developing the calibration plan for NIRSpec. However, certain measurements are extremely difficult to reproduce when not on sky and if using sources unlike real astrophysical targets. To compensate for that, we have also made use of simulations. An instrument performance simulator for NIRSpec was developed by the Centre de Recherche Astrophysique de Lyon (CRAL) for ESA \cite{2012SPIE.8449E..0AJ}. The simulator has been designed as a set of modules that handles different parts of the instrument and its performance: a Fourier optics module that reproduces the optical planes and instrument configurations as well as producing synthetic PSFs in various planes; coordinate transforms that parameterize distortion maps between the major optical planes and allow the computation of the location of any ray traveling through the instrument; a radiometric response module that computes the throughput as a function of wavelength and position in the field of view (FOV); and an exposure simulator to generate synthetic detector readouts. Over the years, the simulator has been updated in accordance with the instrument hardware changes and measurements, ensuring the two remain synchronized. At the moment, simulations are mainly used to understand the path-loss effects that occur as a combination of geometric and diffraction losses. We envision that sky-like simulations will also be used in preparation of commissioning activities. Astrophysical simulations have also been released to the scientific community and are made available at the following location: https://www.cosmos.esa.int/web/jwst/simulations. 

\subsection{On-sky data}
After launch, JWST will go through a commissioning campaign that will ensure, amongst many other things, that NIRSpec is ready for science operations. The scope of the activities covers both functional procedures to activate and assess the instrument's health, and performance tests at the operating temperature and environment to characterize the detectors' response, evaluate the MSA operability, and verify the instrument's alignment, throughput, and image quality \cite{2016SPIE.9904E..44B}. Additionally, this first dataset can also be used to update several areas of the instrument's calibration, an effort that will continue with dedicated activities throughout the science operations phase of the mission.

\section{Detector calibration}
The NIRSpec FPA consists of two HAWAII-2RG HgCdTe detectors. Each detector is paired with a SIDECAR (System for Image Digitization, Enhancement, Control and Retrieval) ASIC (Application-Specific Integrated Circuits), and has a cut-off wavelength of 5.3$\mu$m. The noise performance of the NIRSpec detectors is described elsewhere in these proceedings\cite{Detectors}, and therefore only a high-level summary is presented here.

NIRSpec detectors can be operated using two different readout modes. The \textit{traditional} readout mode is shared between all NIR instruments on JWST, with four outputs being used to readout science pixels following the standard non-destructive 'up-the-ramp' implementation. The new readout mode being offered called \textit{improved reference sampling and subtraction} \cite{2017PASP..129j5003R} (IRS$^2$), takes advantage of a more sophisticated data processing using additional reading of interleaved reference pixel and of the digitized reference output. NIRSpec's detectors can also be used in subarray mode, where only a fraction of pixels is read from the full frame. To increase the dynamical range of the detectors, and driven by the scientific need to support observations of brighter targets, NIRSpec will operate with two different detector electronic gains, with all subarray data being acquired with a higher gain. This allows data to reach the physical saturation of the pixel, instead of being limited by the 16-bit analog to digital converter (A/D) limit (i.e. 65535), as is the case for the full-frame readout modes that are optimised for the observation of faint objects.

The calibration of NIRSpec data at detector level follows standard techniques for data processing of near-IR detectors, namely: \textit{(1)} saturation detection in relation to a measured threshold; \textit{(2)} bias subtraction; \textit{(3)} reference pixel subtraction, which differs significantly according to readout mode; \textit{(4)} non-linearity correction; \textit{(5)} dark current subtraction; and \textit{(6)} slope estimation. The measurements necessary to derive and monitor these corrections were acquired both at DCL level where the detectors were tested prior to being integrated into NIRSpec, and throughout the ISIM and OTIS ground-test campaigns. The data were acquired at a temperature as close as possible to the expected mission operating temperature for the detectors. This ensures that the detector-level testing can be used as a benchmark in the monitoring of several of these corrections, even if they need to be revised once in orbit. 

\section{The NIRSpec instrument geometric model}
NIRSpec will be the first multi-object spectrograph in space offering $\sim$250,000 configurable micro-shutters. The shear number of possible combinations between the available apertures and the disperser was in itself a strong driver to adopt a model-based approach for calibration and spectral extraction. Additionally, the non-repeatability of the grating wheel assembly (GWA) positioning after each movement, precluded the use of a static calibration solution. A parametrization of the NIRSpec optical path that was initially developed to aid the creation of simulations to verify the instruments performance, has been refined and updated with data taken during ground-test campaigns to reflect the instrument \textit{as built}. Below we provide a short description on how the NIRSpec model is derived and used.

\subsection{Spectrograph}
\label{sec:imod_spec}
The NIRSpec spectrograph model has been extensively described in the literature\cite{2016A&A...592A.113D,2016SPIE.9904E..45G}, and therefore we provide here only a short summary for completeness. The model is divided in two major components: \textit{(1)} the parameterization of the coordinate transforms between the main optical planes (IFU-FORE, IFU-POST, collimator, and camera); and \textit{(2)} the geometrical description of their elements (MSA, IFU slicer, GWA, FPA). Optical transforms are modeled using an ideal paraxial system and a fifth-order, 2D polynomial to capture the distortions. The geometry of the MSA and the fixed-slits is described using their relative location to the center of the FOV, rotation angles, and sizes of the apertures. The GWA elements are described by their orientational positioning in the wheel, and according to type, with the gratings defined by the individual groove densities and front surface tilt angles, the prism by the front surface tilt, internal prism angle, and the Seidel model based prescription for its refractive index, and the mirror treated as a simple reflective surface. The geometry of the FPA is described for the two arrays by their absolute position, rotation angle, and size. For the IFU, besides the transforms that were already mentioned, the slicer and each individual slice are also described by their location, rotation angles, and size.

During ground testing, the data necessary for the model optimization, such as imaging, continuum spectra, and spectral lines, were taken using internal calibration lamps. The data were acquired by cycling through the relevant lamps and using different MSA configurations depending on the disperser or the mirror, and used to create the spatial and spectral reference points. After a first manual adjustment to account for any large-scale offsets easily spotted in the data, an automated optimization is carried out that relies on the minimization of the residuals between predicted and measured values from the reference data using a least-squares fit. So far, the instrument model has been verified and updated after each campaign of environmental tests, since these were always preceded by transport and integration of the instrument, and even a hardware upgrade on the instrument itself \cite{2016SPIE.9904E..0DT}. The same procedure will be followed after launch, where the model will be updated to represent the state of the instrument in-orbit. The methodology developed to fit the instrument model has consistently yielded an intrinsic spatial accuracy better than 1/10 of a pixel, and a wavelength calibration showing RMS residuals equivalent to approximately 1/20 of a resolution element\cite{2016A&A...592A.113D}.

\subsection{Grating wheel position sensor}
\label{sec:imod_gwa}
The NIRSpec GWA is a cryogenic wheel mechanism equipped with six dispersion gratings, a prism, and a mirror. Accurate positioning is achieved by use of the selected element, a ball bearing controlled by a cryogenic torque motor and a spring operated ratchet that settles between the index bearings. The finite angular positioning repeatability of the wheel causes small but measurable  displacements of the light beam on the focal plane, precluding a static solution to predict the light-path. To address that, two magneto-resistive position sensors are used to measure the `tip and tilt' displacement of the selected GWA element each time the wheel has rotated into place \cite{2008SPIE.7018E..21W}.
\begin{figure}[h]
\centering
\hspace{\stretch{1}}
\includegraphics[width=0.48\textwidth]{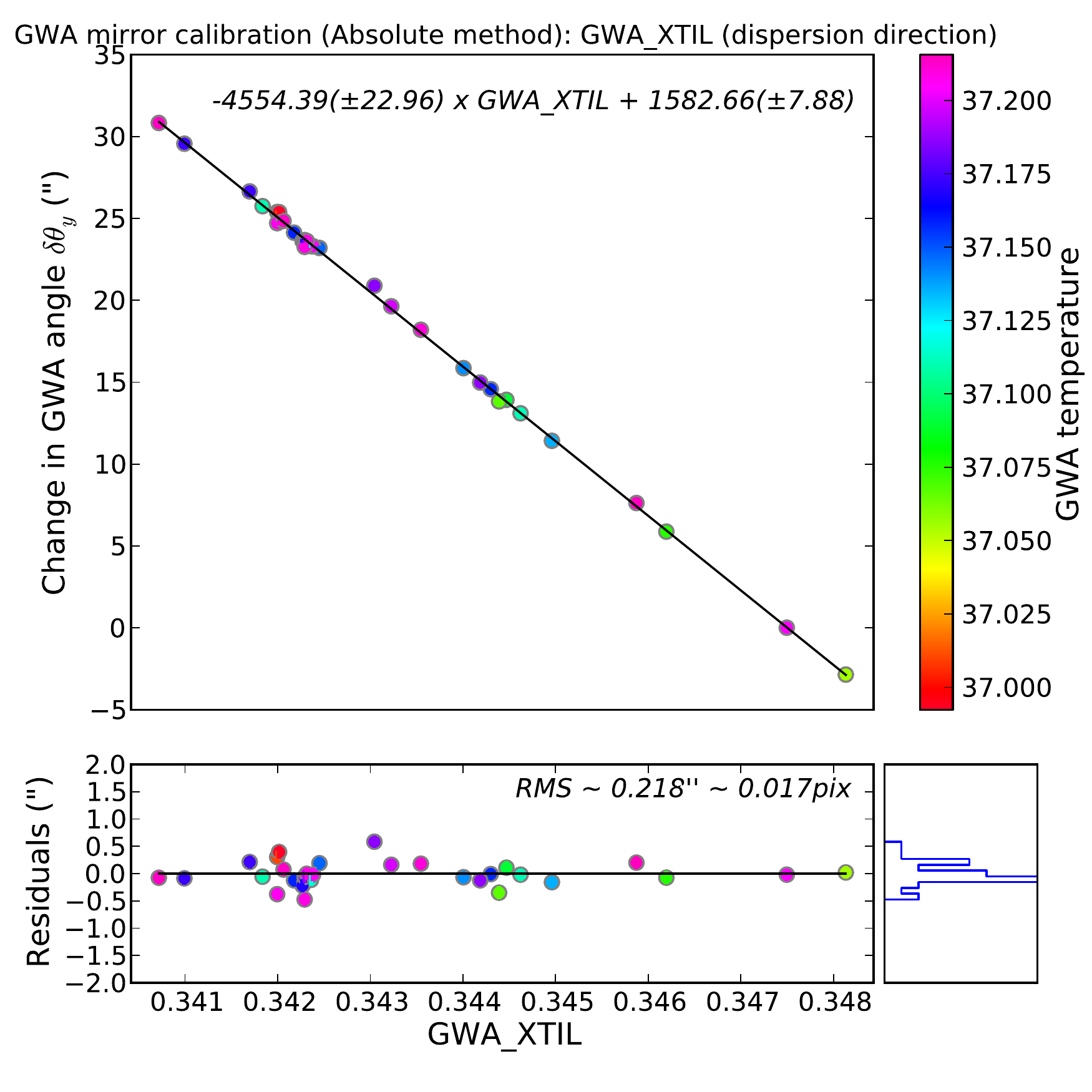}\hspace{\stretch{1}}
\includegraphics[width=0.48\textwidth]{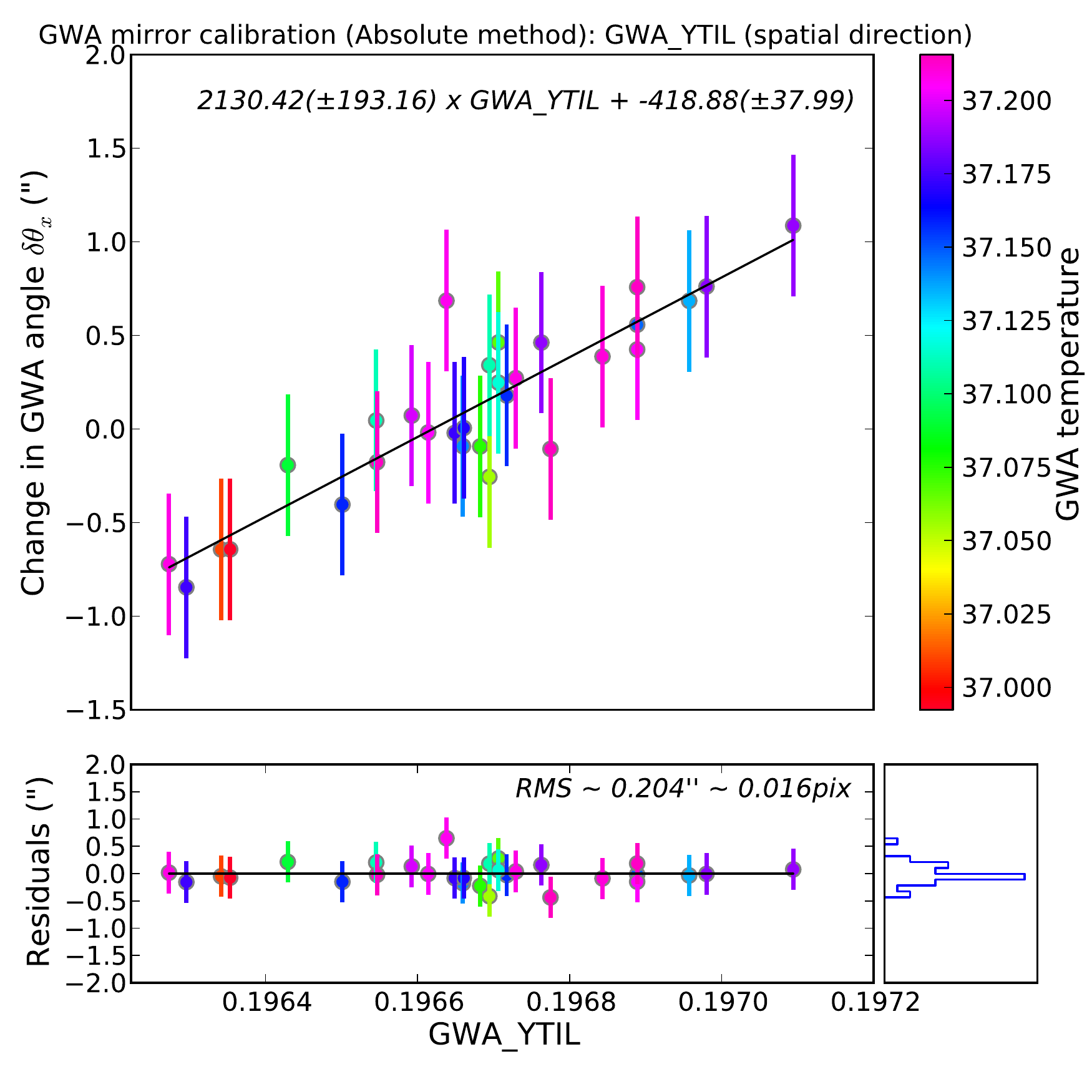}\hspace{\stretch{1}}
\includegraphics[width=\textwidth]{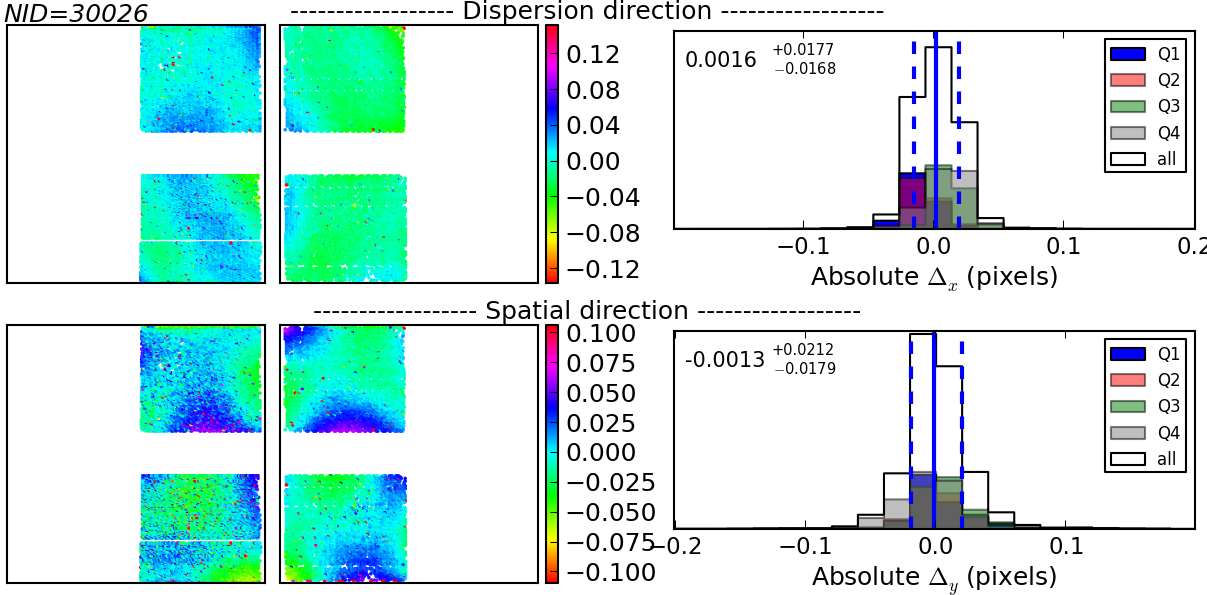}\hspace{\stretch{1}}
\caption{\label{fig:gwa}\textit{Top:} Calibration relations using the derived GWA position for each exposure, and the respective sensor readings. \textit{Bottom:} Difference between simulated and measured shutter centroids at the detectors using the derived GWA angular position, and histograms of the same differences.}
\end{figure}

The calibration of the GWA position sensors aims at finding the relationship between the sensor voltages and the corresponding angular displacement of the wheel, with respect to a reference location, given by the chosen reference point of the spectrograph instrument model. For the dispersers, a number of dedicated exposures are taken with an internal lamp with intermediate wheel movements. Absorption lines in the spectra are used to measure the relative displacement which is translated to arcseconds at the wheel plane. For the mirror, imaging through the MSA configured with a checkerboard-like pattern of opened shutters allows for an analogous measurement of the wheel position based on centroid measurements. The calibration relation between the sensor voltage reading and the measured angular displacement of the wheel has been shown to be linear\cite{2012SPIE.8442E..3GD}. These relations, one per element, are then applied to every NIRSpec exposure to complete the geometrical description of the state of the instrument. Figure~\ref{fig:gwa} shows the calibration relation for the mirror. Since the sensors' response is temperature dependent, and also the mounting of the elements on the GWA is designed to accommodate some movement due to temperature gradients and gravity orientation, the calibration relations will be updated once in-orbit.

\subsection{FORE and OTE optics}
\label{sec:imod_fore}
\begin{figure}[h]
\centering
\hspace{\stretch{1}}
\includegraphics[width=0.49\textwidth]{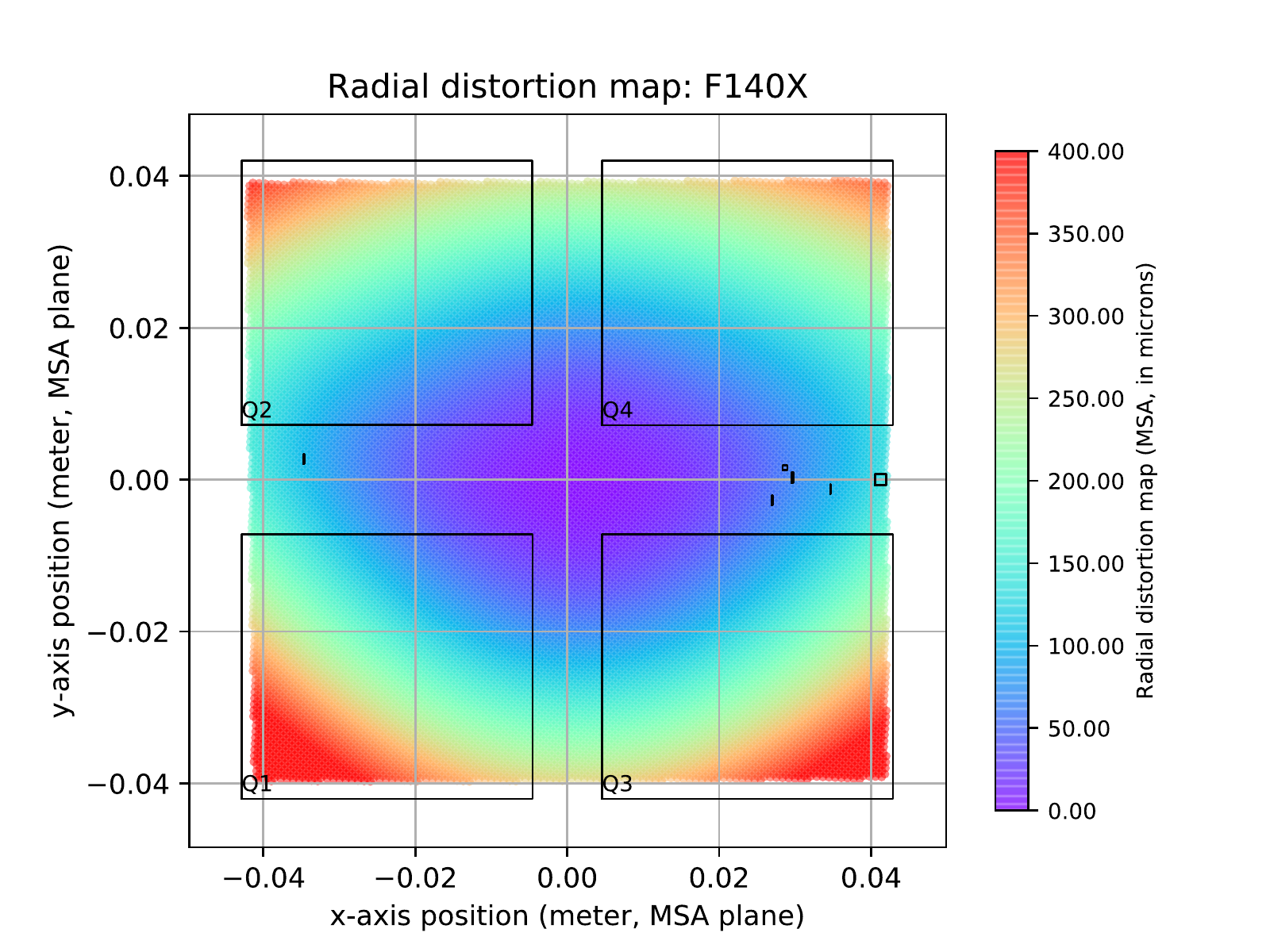}\hspace{\stretch{1}}
\includegraphics[width=0.49\textwidth]{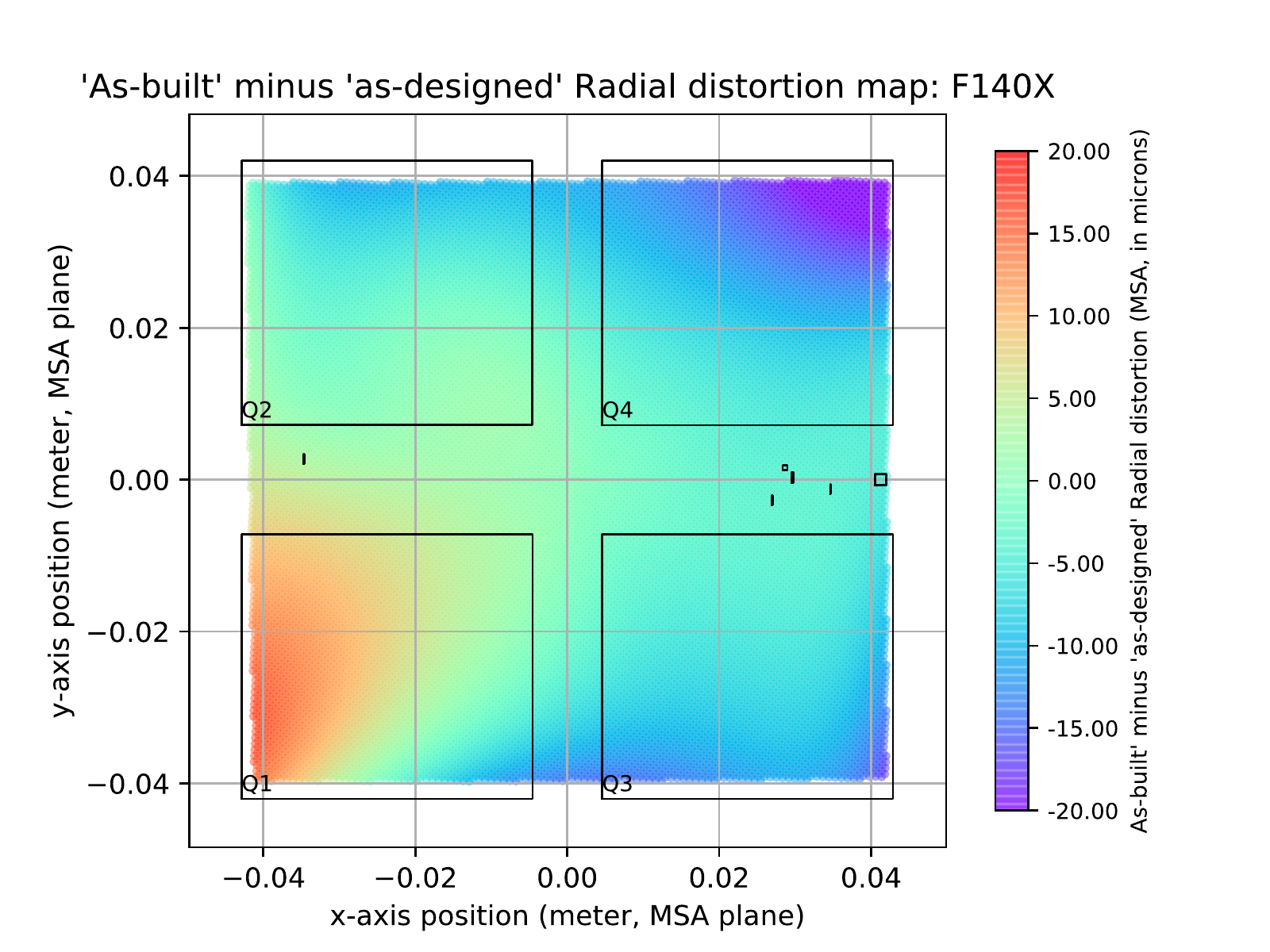}\hspace{\stretch{1}}
\caption{\label{fig:distortion} \textit{Left:} Radial distortion map of the NIRSpec FORE optics for filter F140X derived with data from the instrument level ground test-campaign at IABG facilities. \textit{Right:} Difference between \textit{as designed} and \textit{as built} distortion maps.}
\end{figure}

The science instruments onboard JWST are located behind the Optical Telescope Element (OTE), which provides an image plane (OTEIP). The light entering NIRSpec from the OTEIP is re-imaged by the FORE optics, which includes the Filter Wheel Assembly (FWA) at its pupil plane, onto the aperture plane at the MSA. The existing parametric description of the paraxial terms and distortion of the FORE optics is based on the \textit{as-designed} model. In the first ground-test campaign however, a pinhole-mask was used to acquire characterization data through each filter. The images of this grid of small-diameter pinholes measured on the detectors can be used to track how an image in the OTEIP plane is projected on the FPA plane. Using the spectrograph instrument model description, the location of the pinholes on the MSA plane can be calculated. The FORE optics model is then derived by minimizing the residuals between the model predicted and measured values of the pinholes centroids using a least-squares fit. A comparison between the distortion maps measured for the \textit{as built} instrument to the designed, showed the differences to be at the $\sim$$\pm$5\% level. Figure~\ref{fig:distortion} shows the radial distortion map for one of the NIRSpec filters. In subsequent ground-test campaigns, NIRSpec was already integrated with the other instruments, and there was no possibility to repeat this test. These results are however important in confirming the validity of the current FORE optics model. Once in-orbit, the final alignment and orientation of NIRSpec on the JWST focal plane, as well as the FORE optics distortion will be evaluated and updated using observations of a pre-selected astrometric reference field for JWST. Imaging data will be acquired through the seven filters with the MSA configured to have all shutters open. The positions of thousands of stars that will fall in the NIRSpec FOV will be measured and matched to their known coordinates on the GAIA reference frame, to serve as the reference data to be used in the fit.

\subsection{Wavelength zero-point correction for off-centred sources}
\label{sec:imod_wlzp}
NIRSpec operates mostly in the diffraction-limited regime and over a large wavelength range (almost a factor of ten in wavelength). As a consequence, the size of the point-spread function (PSF) varies considerably and it was not possible to select aperture sizes that would match the width of the PSF at all wavelengths. In addition, NIRSpec was optimised for observations of faint objects, so in order to maximize the sensitivity, the choice was made to use slightly oversized apertures with a typical width of 200 milli-arcseconds for the micro-shutters and the standard slits. As a consequence, strong intensity gradients can be present across the aperture and NIRSpec is susceptible to the \textit{slit-effect} where an imprint of the intensity distribution of the object across the slit will be present in the spectral PSF. To first order, this means that for compact sources and in particular point sources the position of the barycentre of the spectral PSF will depend on the centering of the source in the aperture. It will also be different from the one measured in the uniform illumination case, i.e. from the one used for the wavelength calibration of NIRSpec. A wavelength offset of a fraction of a pixel will therefore be present that needs to be corrected.

As part of the ground testing of the NIRSpec instrument, unblended spectrally unresolved emission-line sources were not available with a point-source illumination configuration, therefore this calibration currently relies on modeling. For a grid of wavelengths and source positions within a given aperture, for a point source, a coherent Fourier transform propagation through the instrument was computed, while the uniform illumination case relied on the convolution by the point-spread function of various optical modules. The corrections were then constructed by computing the relative barycenter offset between the point-source and uniform cases for the same grid. Once in orbit, we plan to use commissioning data acquired for other activities to correlate the model with observations at the available wavelengths and source positions.

\subsection{Applications}
\label{sec:imod_app}

The NIRSpec instrument model is the backbone of several aspects of NIRSpec science operations. It is at the core of the algorithm developed to plan the MOS observations, where the expected location of science targets on the MSA plane needs to be known in order to derive an optimized mask of open-shutters that maximizes the number of targets observed. For that, the coordinate transforms of the FORE optics and accurate knowledge of the MSA geometry, together with the operability status of the MSA\cite{MSAStatus} are needed. The final step in this process is the fully automated on-board target-acquisition. By using NIRSpec in its imaging mode, a set of images are taken at the beginning of an observation, a centroid algorithm measures the locations of the targets, after which the instrument model is used to convert the coordinates so that they can be compared to the desired location, and the corrective slew can be derived.

For science data processing, an algorithm was developed for an automated spectral extraction that derives the 2D spectral coordinates (wavelength and spatial coordinate in the aperture) using the instrument model\cite{2016A&A...592A.113D,2016SPIE.9904E..45G}. This ability to extract wavelength-calibrated spectra from any of the $\sim$250,000 NIRSpec apertures (it is also applicable to data acquired with the fixed-slits or the IFU) using this model-based approach greatly improves the efficiency with which NIRSpec data can be processed. In particular, the conventional and time costly approach of acquiring dedicated calibration exposures for each observation and carrying separate empirical calibration for each individual aperture is circumvented.

Finally, the instrument model is also used as a tool to derive other parameters such as the pixel area used for the calibration of science data, or the optimized dithering patterns and sizes to improve sampling. 

\section{The NIRSpec radiometric calibration}
The radiometric calibration of the NIRSpec instrument is based on a fairly standard combination of flat-fielding and spectro-photometric correction steps. Some of the corrections, for example parts of the flatfielding step, can be derived using internal lamps and data acquired on the ground (e.g., at detector testing level). Other calibrations can be modeled but will ultimately have to be measured on-sky for each NIRSpec mode. In the wavelength range and sensitivity of NIRSpec, the available flux standards for spectrophotometric calibration are stars, and therefore the radiometric calibration of NIRSpec will be anchored on a point source centered in the respective aperture. As described below, corrections are then derived to account for either non-centred point or extended sources.

\subsection{Flatfield}
\label{sec:flatfield}
The flatfield correction addresses the variations in throughput as a function of the location of a source within NIRSpec's FOV and wavelength. By design of the calibration plan, it was also convenient to incorporate the detector's response in this correction. As in any spectrograph, the flat-field correction is conceptually more challenging than for an imager, as each disperser introduces a different wavelength dependence of the instrument throughput. In the case of NIRSpec's MOS mode, this difficulty is compounded by the fact that the same detector pixel can be illuminated by many different wavelengths, depending on which shutter was opened to illuminate the pixel. This also has the consequence that flatfielding of NIRSpec data can only occur after the wavelength falling onto a given pixel has been determined. We summarize in this section the approach adopted to deal with these complexities\cite{2016SPIE.9904E..46R}.

The NIRSpec flatfield correction has been split according to three different parts of the optical path: detectors, spectrograph, and FORE optics (including the telescope). The driving factors of this design were manifold. The existence of a calibration lamp assembly inside NIRSpec containing five continuum sources optimized for measuring the instrument's response provides a stable and suitable calibrator for the spectrograph. The detector performance characterization obtained prior to their integration on to NIRSpec allowed a flatfield correction to be derived from a uniform illumination of their entire area, which is not possible once the MSA is in the optical path. The on-sky measurements are then needed only to characterize the FORE optics and OTE flatfield variations, for which observations of a point-source at each aperture are sufficient (except for the MOS mode, where a few measurements per MSA quadrant are needed). Additionally, for each of the flatfield parts, the correction is divided between the spatially-independent spectral variations, which are traced by a fast-varying vector, and the spatially-dependent variations that can be derived with a sparse spectral sampling.

\subsubsection{Detector (D-flat)}
\label{sec:dflt}
The detector component of the NIRSpec flatfield (D-flat) corrects for the wavelength-dependent responsive quantum efficiency of any given pixel. It is derived using measurements taken with monochromatic sources that probed the full operational spectral range of NIRSpec at 39 different wavelengths. As already mentioned before, this was achieved prior to integration into the instrument. Additionally, the variations caused by the anti-reflection coating were measured at a higher spectral resolution by the manufacturer (Teledyne Imaging Systems), and those are also taken into account when deriving the D-flat.

\subsubsection{Spectrograph (S-Flat)}
\label{sec:sflt}
The spectrograph component of the flatfield corrects throughput variations that occur between the aperture focal plane (MSA plane for micro-shutters and fixed-slits, slicer plane for IFU), and the detectors. The S-Flat corrections are determined via exposures of the internal calibration flat lamps. In the cases of the fixed-slits and IFU, for a given instrument configuration (filter and grating), the incident light on any given pixel will always have the same wavelength, apart from small variations due to the inherent uncertainty in GWA position. For the MOS mode, the correspondence between wavelength and pixel further depends on which micro-shutters are open. With ~250,000 shutters, there are simply too many possibilities for all shutters to have their flatfield measured with all nine standard instrument configurations. The calibration relies instead on acquiring data on a subset of selected shutters, from which the S-flat for any given shutter can be interpolated\cite{2016SPIE.9904E..46R}.

\subsubsection{Fore Optics (F-Flat)}
\label{sec:fflt}
The F-Flat includes the relative throughput losses of all the reflections along the JWST optical telescope element (OTE) and the NIRSpec Fore Optics. This correction can only be derived on-sky, and given it is a critical step towards verifying the instrument's health, and towards validating the assumptions on throughput performance, it will be measured during commissioning. Observations of a spectrophotometric standard star across the NIRSpec FOV will be used to quantify any field- and wavelength-dependent effects caused by the Optical Telescope Element (OTE) and the NIRSpec Fore Optics, neither of which can be traced with the internal CAA lamps. The F-Flat also includes the transmission curve of the selected filter located in the FWA. Due to NIRSpec's design, light from the internal calibration assembly does not pass along the optical path probed by the F-Flat.

\subsection{Pathloss corrections}
\label{sec:pthlbars}
The pathloss of light going through the NIRSpec spectrograph is caused mainly by geometric and diffraction losses. Additionally, for uniform-type sources observed through the micro-shutters, although the geometric losses are not relevant, the presence of the micro-shutter bars will also result in flux losses.\\

\subsubsection{Geometric and diffraction losses}
Geometric losses (also commonly referred to as slit losses), occur when the reduced size of the aperture restricts the incoming flux collected from a target by the telescope. In NIRSpec, at longer wavelengths, geometric losses increase as a result of the broadening of the PSF due to diffraction (OTE and NIRSpec's FORE optics)\cite{2007SPIE.6692E..0NT}. The effect of geometric losses for NIRSpec fixed-slit and MOS observations also depends on the source location within the aperture. For data taken with the IFU, geometric losses can be ignored since the light falling outside a slice is not lost but collected by the neighbouring slices. The other contributor to pathlosses are diffraction losses. The truncation that takes place in the image plane (fixed-slit and micro-shutter apertures, or slicer for the IFU) causes a widening of the illumination to the extent that the beam is larger than the surface of the disperser in the GWA, resulting in a partial loss of light. The pathloss correction addresses both types of losses together, since quite often it is not possible to disentangle them, especially when using on-sky observations. The correction depends on several parameters, such as the aperture width, the type of object (point-source or extended), for point-sources their position within the aperture, and wavelength.

\begin{figure}[h]
\centering
\hspace{\stretch{1}}
\includegraphics[height=0.5\textwidth]{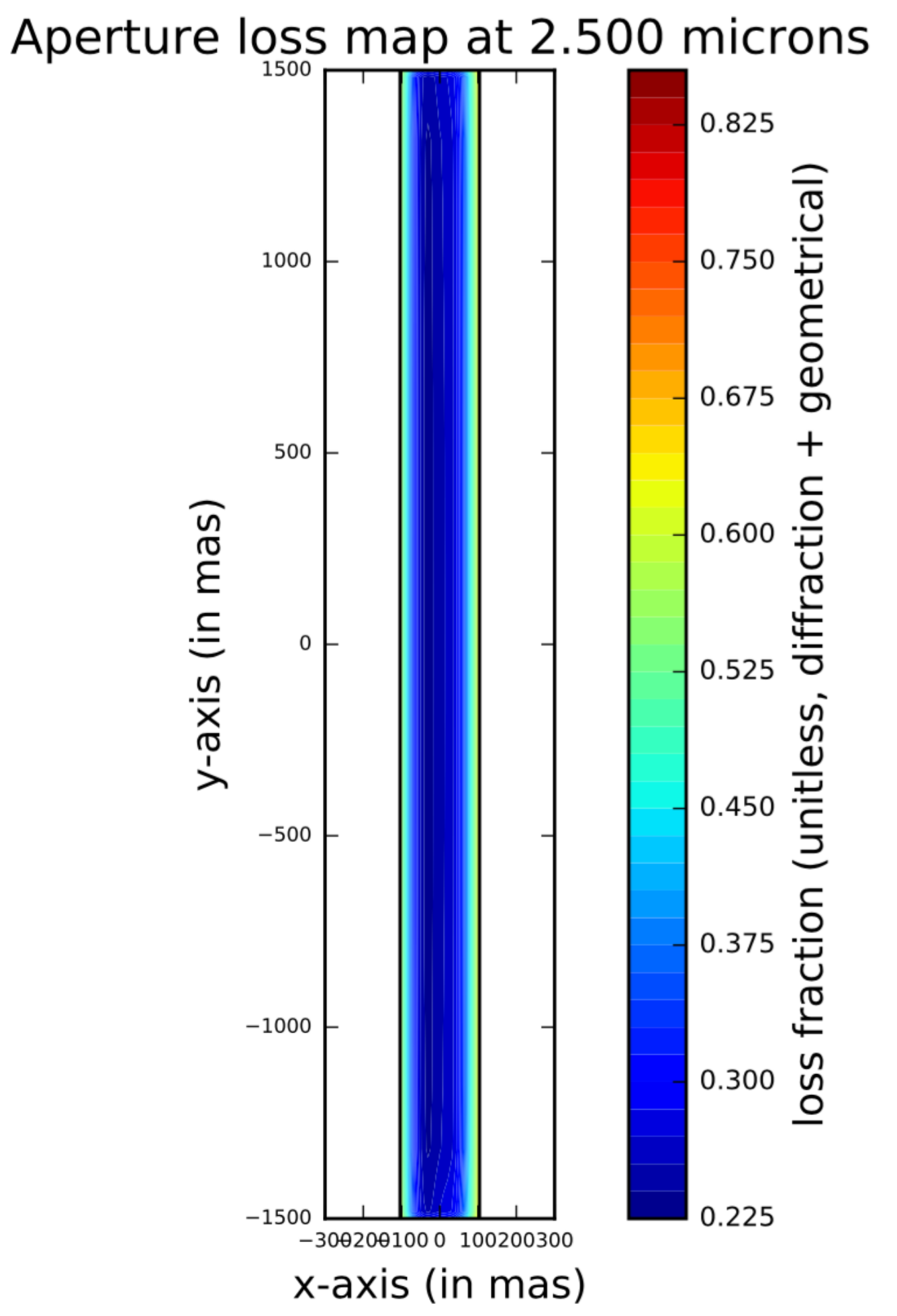}\hspace{\stretch{1}}
\includegraphics[height=0.5\textwidth]{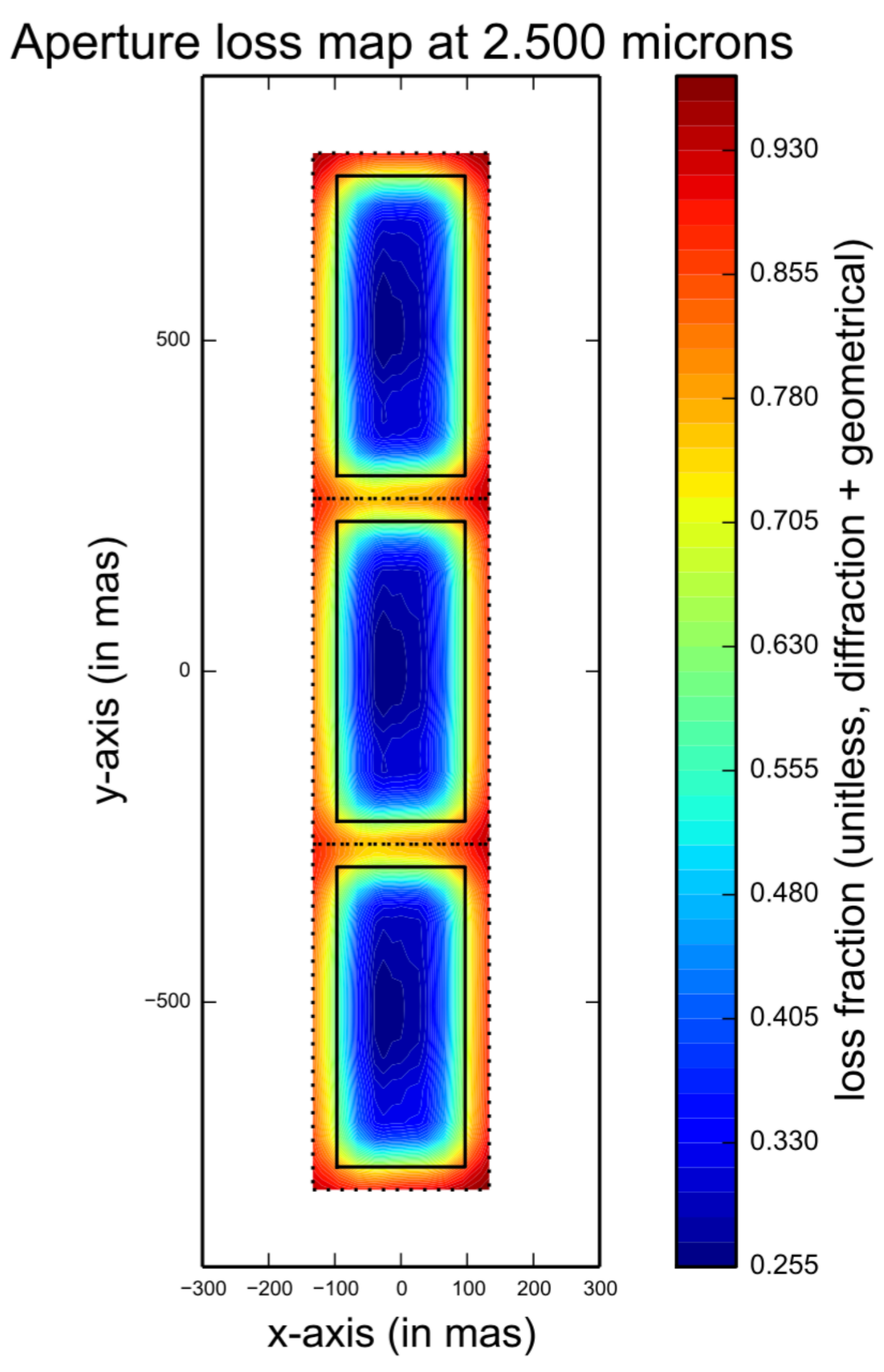}\hspace{\stretch{1}}
\caption{\label{fig:pathloss} Example of modeled pathloss correction maps for a point source at 2.5~$\mu$m for a slit \textit{(left)} and a MOS1x3 mini-slit \textit{(right)} cases. Overlayed solid lines depict aperture sizes, while dashed lines represent the pitch size for the MSA shutters.}
\end{figure}

The current strategy to study this effect on the ground is to use a Fourier-optics model of the OTE and NIRSpec instrument to compute point-spread functions and pupil illumination in various instrument planes. The model starts from the pupil of the telescope to the plane of the MSA (passing through the plane of the FWA). This includes the OTE and FORE-optics of NIRSpec, plus the IFU fore-optics for the IFS mode. It then goes from the plane of the MSA (or slicer) to the plane of the disperser, corresponding to the collimator, plus the IFU post-optics for the IFS mode. The third component corresponds to the camera for all modes, and it applies the computation from the plane of the disperser to the plane of the detectors. Using these computations, the slit and diffraction losses can be estimated. For point-sources observed with the fixed-slits or MOS mode, the model-based corrections were simply combined and converted into efficiencies relative to a perfectly centred target at each wavelength. Figure~\ref{fig:pathloss} shows two examples of the modeled corrections. The model-based correction for point-like sources observed with the IFU needs additionally to take into account the neighbouring slices when computing the intensity-weighted sum of the diffraction losses. The pathloss correction for uniform sources also includes the correction required to go from a centred point source, which is the reference of NIRSpec's radiometric calibration, to a uniform illumination source. The current model assumes a fully non-coherent illumination with no diffraction effects, resulting in that the correction for a uniform source corresponds to the inverse of the pathlosses correction for a centred source. Improvements to this model are currently being developed. 

Once in-orbit, we plan to make use of data taken during commissioning to start characterizing this effect with astrophysical sources, for example, by using data from the activity that will verify the NIRSpec throughput as a function of wavelength and field position. For observations taken with the fixed-slits and the IFU, the plan includes dithering a spectrophotometric standard star within the aperture, probing the default locations where most science targets will be placed. The empirically derived correction will be very sparse in terms of coverage of the aperture area. Additional measurements will then be acquired throughout the lifetime of the mission. We envision however, that if observations concur with the predictions of our model, a hybrid methodology could be adopted to provide the most accurate pathloss corrections. For MOS observations, the desire to increase multiplexing will come at the cost of relaxing centering constrains, and we expect that most NIRSpec MOS targets that require accurate flux calibration will need to be corrected for pathlosses. A dedicated commissioning activity is planned to characterize the wavelength-dependent throughput variations as a function of position within a typical NIRSpec micro-shutter. The observations will target a field with 20 stars distributed across the MSA FOV. A customized MSA configuration will place at least one of these stars in the center of a three-shutter slitlet, while the remaining ones will be observed at different positions within the central shutter of their respective slitlet. This observing strategy is then repeated approximately 20 times, until each of the ~20 stars has been observed through the best centering of a shutter, and several times at other locations within the slitlet. The measured pathloss map is then constructed by combining the relative difference in flux for each star between spectra obtained at different locations within the shutter with respect to the central measurement. This approach assumes that all shutters behave identically, given that measuring $\sim$250,000 shutters individually is an insurmountable task. As for the other NIRSpec modes, a comparison between model-based and empirically derived corrections will be conducted.

\subsubsection{MSA bar-shadow correction}
\label{sec:bars}

When using NIRSpec in MOS mode, adjacent micro-shutters are opened to emulate a \textit{slitlet}, which is however not continuous but interrupted by the presence of bars between the shutters. These bars will mask part of the sources, whether they are point or uniformly extended sources such as the sky background, and this effect needs to be calibrated. For point sources, no attempt is made to correct for the impact of the bars on the spatial profile of the source on the detectors but, by design of NIRSpec's radiometric calibration scheme, the associated loss of light is accounted for in the pathloss correction that was described in the previous section so that when collapsing the data along the spatial direction the resulting 1D spectrum is correctly calibrated. For uniform sources, we also correct the spatial profile of the sources in a special step called the MSA bar-shadow correction. Assuming uniform illumination, there are two ways in which the bars affect the flux for a given pixel. On the one hand, the size of the projection of the shadow of the bars will vary with wavelength since it is the convolution of the shape of the bars by the PSF of the spectrograph. On the other hand, there is an aliasing effect which is a function of how the sampling is performed. The fact that the spectra are not aligned with the detector rows and columns translates into a rapid evolution of the aliasing effect. A correction is therefore dependent on the profile of the optical image of the bars along the spatial direction on the detectors, any cross talk between neighbouring pixels, and the actual sampling of the image of the bar by a given pixel.

To derive this calibration, work is on-going to compare corrections derived either empirically or through modeling. For the modeling approach, the shape of the optical image of the bars is taken to be a function of the width of the micro-shutter bars along the spatial direction. It is assumed to be identical for all micro-shutters and its value is given by the difference between the pitch and aperture size. The point-spread function (PSF) of the spectrograph varies with wavelength and with the path through the spectrograph. It is also assumed that the dominant effect is the dependence with wavelength, and other variations across the spectrograph are neglected. The inter-pixel cross-talk is assumed to be the same for all pixels. The image of the micro-shutter bars or edges on the detector was computed using non-coherent propagation through the spectrograph (i.e., convolution with a model spectrograph PSF). It was then collapsed along the spectral direction to get a 1D profile and convolved with a door function to simulate the sampling by a pixel, yielding a 1D profile covering the dependence with pixel sampling. This was repeated for a grid of wavelengths, allowing the generation of a 2D image correction. To derive the same correction empirically for a uniform illumination source, NIRSpec data taken with internal calibration lamps can be used to create the pixel-sampled intensity profile of the shadow, which in turn can be directly compared to the one modeled.

\subsection{Radiometric correction factor}
\label{sec:radm}

For science data, a final wavelength-dependent radiometric calibration step is needed, which includes any residual correction not anticipated and not captured by the preceding radiometric corrections. As stated before, the on-ground NIRSpec radiometric calibration is limited as some of the necessary measurements can only be acquired once in orbit, which is the case for this radiometric correction factor. After launch, we anticipate that data acquired during commissioning for the primary goal of measuring the NIRSpec throughput as a function of wavelength and field position, can be used to derive this final radiometric correction. As the in-orbit calibration of NIRSpec progresses during the mission, different spectrophotometric standard stars will be observed that may uncover some inaccuracies in the radiometric calibration via discrepancies in their observed spectra. The radiometric correction would need to be updated to account for such wavelength-dependent biases in the radiometric calibration.

\section{Science calibration through operational strategies}
\subsection{Background contributions to NIRSpec data}
\begin{figure}[h]
\centering
\hspace{\stretch{1}}
\includegraphics[width=\textwidth]{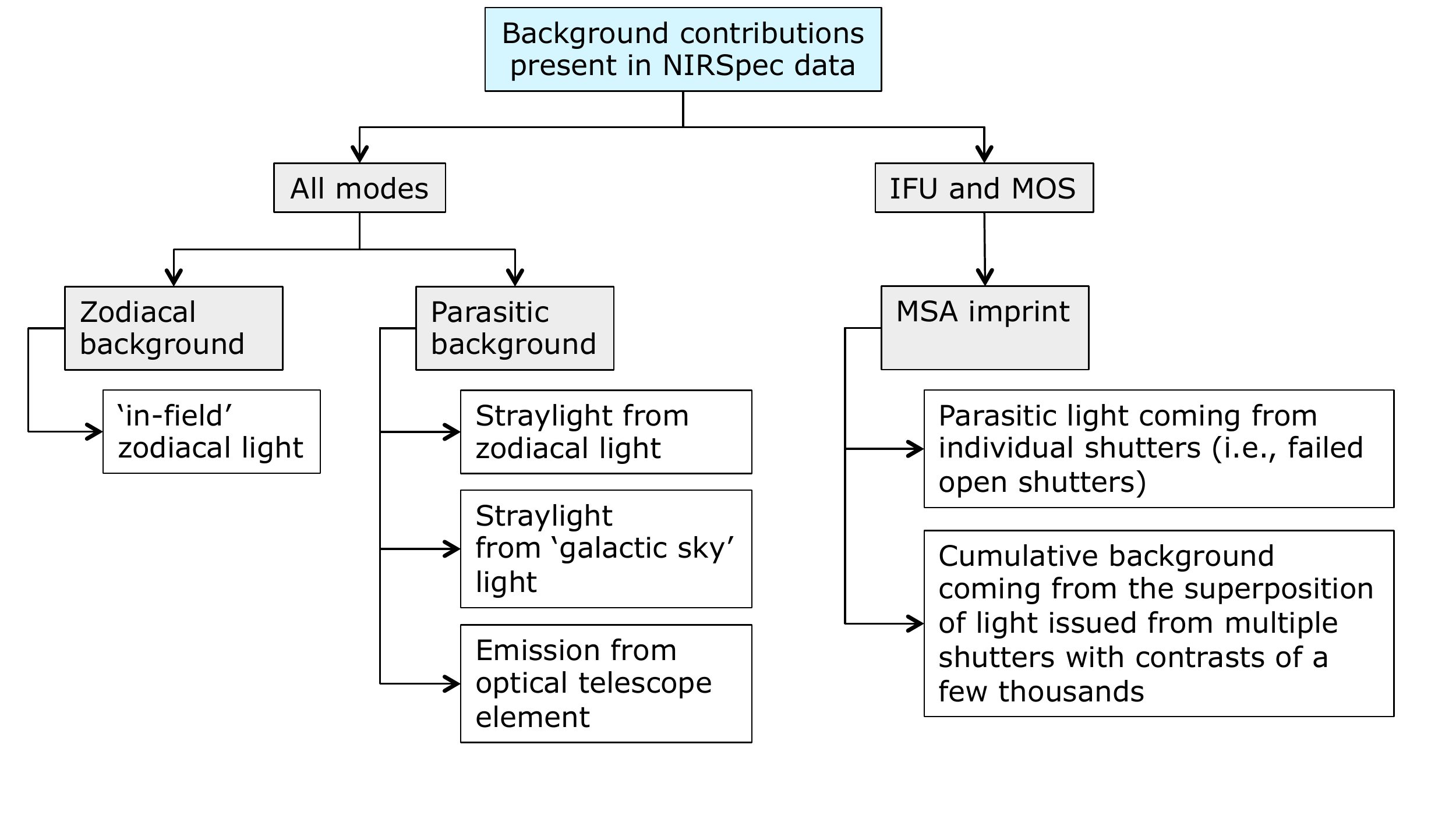}\hspace{\stretch{1}}
\caption{\label{fig:background} Diagram of NIRSpec background contributions.}
\end{figure}

The JWST background contributions that will be present and affect all NIRSpec spectroscopic observations, independently of science or mode, are the zodiacal and parasitic backgrounds. The zodiacal background corresponds to the contribution of the `in-field' zodiacal light. The parasitic background includes two straylight components corresponding to straylight signal scattered into the NIRSpec FOV via rogue paths and, to a much lesser extent for NIRSpec's operating wavelength range, the thermal self-emission from the optical telescope element. Additionally, given that the (closed) MSA micro-shutters are not fully opaque, the imprint of the sky signal leaking through the MSA is visible as a parasitic background both in observations with the MSA and the IFU since they share the same area on the detector. This is caused by two different mechanisms: on the one hand background coming from individual shutters (i.e., failed open shutters), and, on the other hand, the cumulative background coming from the superposition of light originating from multiple shutters with contrasts of a few thousands. The first can only be corrected by a dedicated exposure, while the second can be mitigated either by a dedicated exposure or a model. Figure~\ref{fig:background} summarizes these contributions.\\

\subsection{Background subtraction}
\label{sec:msbg}
The background subtraction of NIRSpec data can be achieved using two standard approaches for near-IR spectroscopy: one where the removed sky background is measured at the same location within the instrument FOV, but at a different location on the sky, the other being the case where the background is determined from sky pixels and rebinned to a common wavelength scale to be appropriately subtracted from the two-dimensional spectra of the target. The direct background subtraction entails a one-to-one subtraction between count-rate exposures, where the flux of each pixel in a given exposure gets subtracted by the flux of the exact same pixel in the other exposure. Examples of this strategy are nodding sequences, or obtaining a dedicated background exposure at another position on sky. The only consideration that is specific to NIRSpec is that this observing strategy needs to ensure that the GWA has not moved between exposures, due to its limited repositioning accuracy, as discussed in Sect.~4.2. 

The other background subtraction strategy consists of the creation of a master background from predefined sky pixels all the way to a calibrated 1D spectrum, which then gets expanded into the 2D unrectified pixel and wavelength grid of the science target. For uniform, extended sources, the pseudo-background is simply subtracted from the calibrated 2D spectrum. For point sources, however, since the source may have been off-centred and subjected to the wavelength zero-point correction, all radiometric calibrations derived at the wavelength and position of the science target need to be removed from the pseudo-background spectrum before it is subtracted. For both methods, their accuracy depends strongly on the assumption that the background present at the science target location is comparable to that measured at a different location. Such assessment is always specific to each science-case. \\

\subsection{MSA-leakage subtraction}
\label{sec:leak}
To correct this specific type of MSA imprint, it is possible to acquire a dedicated exposure taken with the IFU in its closed position and all the micro-shutters commanded closed. Providing that the GWA has not moved between exposures, the MSA leakage can be subtracted directly from the science, pixel-by-pixel. The need to correct for the parasitic contribution of the MSA-leakage depends entirely on the science case, and the decision process involves a trade-off between signal-to-noise (a correction using a dedicated exposure could have a penalty in S/N, since any arithmetic operation of image frames tends to add noise), time needed for obtaining the exposure, and the required level of accuracy of the correction. For certain applications, the use of a model of the MSA-leakage\cite{MSALeakage} may be preferable. The model could be scaled directly to the target exposure by measuring the level of MSA-leakage contribution in the detector areas not being exposed, for example, areas without spectra for MOS observations or the pixels between the slices of the IFU for observations taken in that mode.

\section{The calibration process of NIRSpec science data}
\begin{figure}[h]
\centering
\hspace{\stretch{1}}
\includegraphics[width=0.8\textwidth]{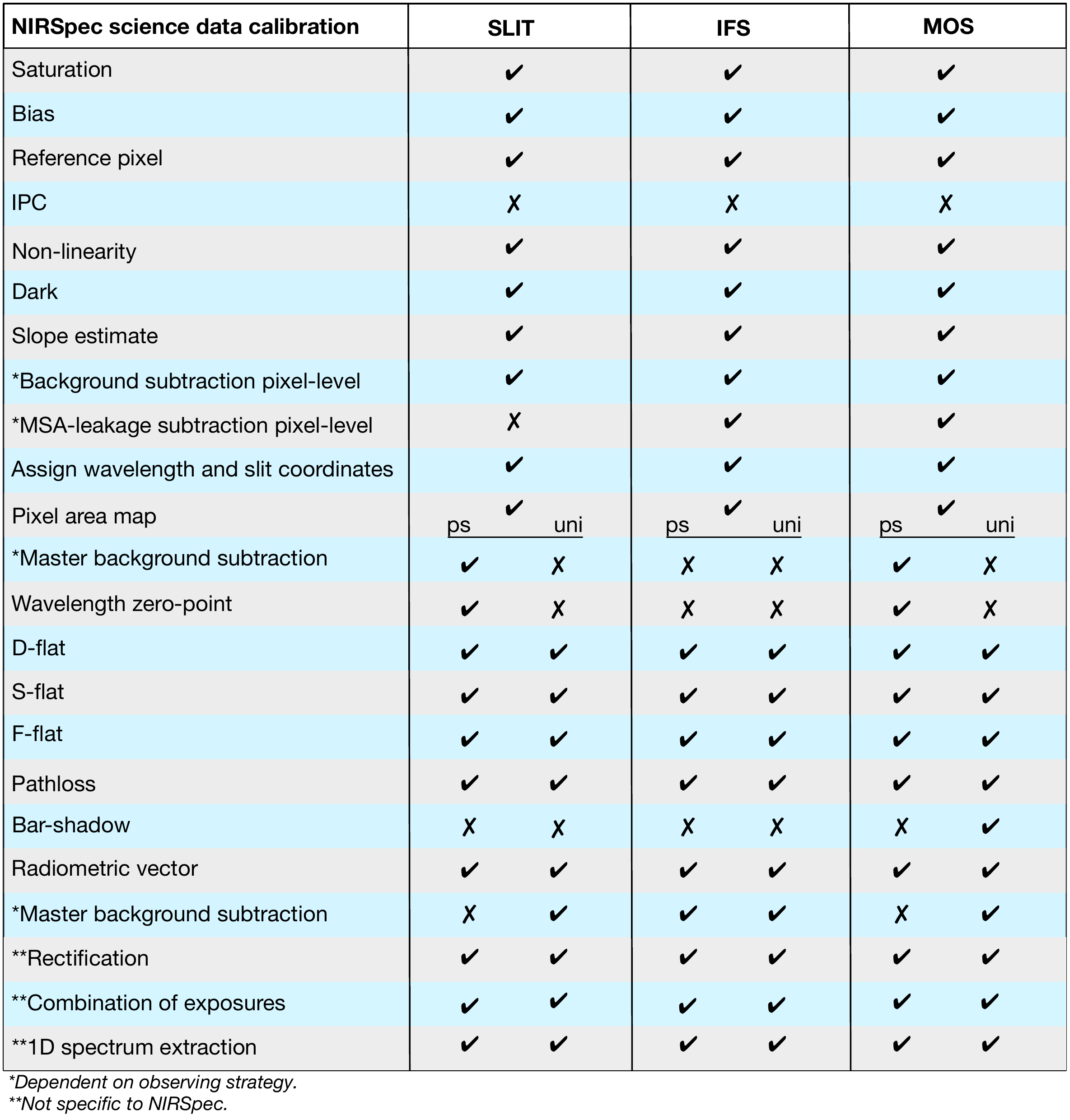}\hspace{\stretch{1}}
\caption{\label{fig:summaryTable} Diagram of the calibration process of NIRSpec science data for its three spectroscopic modes, for a point (ps) or uniformly extended (uni) source.}
\end{figure}

Figure~\ref{fig:summaryTable} provides an overview of all the steps involved in the calibration process of NIRSpec science data. Although most of them apply to all the modes (albeit at times with different reference files and algorithms), some are specific to one or two modes only. For JWST, the Space Telescope Science Institute (STScI) is responsible for the development of the operations ground segment, including the calibration pipeline for all instruments. For the NIRSpec instrument, this is based on the calibration plan described in this proceedings, and its associated algorithms and reference files.

\section{Conclusion}
We present a high-level summary of the calibration scheme developed for the NIRSpec instrument on-board JWST, which is based on modeling and extensive characterization of the instrument over several ground test campaigns. Once in-orbit, dedicated observations will allow us to verify the accuracy of the current scheme, and most importantly to anchor it on astrophysical reference systems.

\acknowledgments 
 JWST is a joint mission between NASA, ESA and the Canadian Space Agency (CSA). NIRSpec was designed and built for the European Space Agency (ESA) by Airbus Defense and Space GmbH in Ottobrunn (Germany), with the two detectors and the micro-shutter array provided by NASA Goddard Space Flight Center.

\bibliography{report} 
\bibliographystyle{spiebib} 

\end{document}